# Structural Phase Transition and Dielectric Relaxation in $Pb(Zn_{1/3}Nb_{2/3})O_3$ Single Crystals

Y.-H. Bing, A. A. Bokov, and Z.-G. Ye

*Department of Chemistry, Simon Fraser University, 8888 University Drive*
*Burnaby, BC, V5A 1S6, Canada*
and

B. Noheda and G. Shirane
*Physics Department, Brookhaven National Laboratory, Upton, New York 11973-500, USA*

The structure and the dielectric properties of $Pb(Zn_{1/3}Nb_{2/3})O_3$ (PZN) crystal have been investigated by means of high-resolution synchrotron x-ray diffraction (with an x-ray energy of 32 keV) and dielectric spectroscopy (in the frequency range of 100 Hz – 1 MHz). At high temperatures, the PZN crystal exhibits a cubic symmetry and polar nanoregions inherent to relaxor ferroelectrics are present, as evidenced by the single (222) Bragg peak and by the noticeable tails at the basis of the peak. At low temperatures, in addition to the well-known rhombohedral phase, another low-symmetry, probably ferroelectric, phase is found. The two phases coexist in the form of mesoscopic domains. The para- to ferroelectric phase transition is diffused and observed between 325 and 390 K, where the concentration of the low-temperature phases gradually increases and the cubic phase disappears upon cooling. However, no dielectric anomalies can be detected in the temperature range of diffuse phase transition. The temperature dependence of the dielectric constant show the maximum at higher temperature ($T_m$ = 417 – 429 K, depending on frequency) with the typical relaxor dispersion at $T < T_m$ and the frequency dependence of $T_m$ fitted to the Vogel-Fulcher relation. Application of an electric field upon cooling from the cubic phase or poling the crystal in the ferroelectric phase gives rise to a sharp anomaly of the dielectric constant at $T \approx 390$ K and diminishes greatly the dispersion at lower temperatures, but the dielectric relaxation process around $T_m$ remains qualitatively unchanged. The results are discussed in the framework of the present models of relaxors and in comparison with the prototypical relaxor ferroelectric $Pb(Mg_{1/3}Nb_{2/3})O_3$.

## I. INTRODUCTION

$Pb(Zn_{1/3}Nb_{2/3})O_3$ (PZN) and $Pb(Mg_{1/3}Nb_{2/3})O_3$ (PMN) are two prototypical relaxor ferroelectric materials with complex perovskite structure, in which the off-valent $Zn^{2+}$ (or $Mg^{2+}$) and $Nb^{5+}$ ions occupying the B-sites are primarily disordered.[1] Research on relaxor ferroelectrics and related materials has undergone an accelerated growth in the last few years both in fundamental understanding of the structure and physical properties and in practical applications. This is partly due to the excellent piezoelectric properties discovered in the single crystals of the solid solutions between PZN (or PMN) and ferroelectric $PbTiO_3$, which point to the next generation of electromechanical transducer applications.[2,3] Recent neutron scattering studies have identified a ferroelectric soft mode in PMN at 1100 K that becomes overdamped below the Burns temperature $T_d \approx 620$ K (i.e. the temperature at which polar nanoregions, PNRs, begin to appear), suggesting a direct connection between the soft mode and the PNRs.[4] More interestingly, at lower temperature the soft mode in PMN reappears close to $T_C = 213$ K,[5] temperature at which the electric field-induced polarization vanishes spontaneously upon zero field heating,[6] and a peak in the temperature dependence of the hypersonic damping appears.[7] To interpret the measured intensities of the diffuse scattering in PMN in accordance with the concept of ferroelectric soft mode, Hirota *et al.*[8] have proposed and demonstrated the validity of a phase-shifted condensed soft mode model of the PNRs. This model suggests the displacement of PNRs along their polar directions relative to the surrounding cubic matrix (H-shift). Therefore, the phonon dynamics clearly indicates the ferroelectric nature of the relaxor PMN, although the average structure of the system remains cubic and optically isotropic.

Application of an electric field along <111> direction can induce a long-range (single domain) ferroelectric phase in PMN, with the development of a polar rhombohedral *R*3m phase associated with switchable polarization and birefringent macro domains.[6,9,10]

By means of dielectric spectroscopy, Bokov and Ye[11-13] have discovered a "universal" relaxor dispersion in PMN and related materials, and showed that it is an important common property of relaxor ferroelectrics. The universal relaxor polarization is described by a microscopic model of 'soft' polar nanoregions with unit cells that can freely choose several different directions, while the direction of the total moment of the nanoregion remains the same.[13] Such approach makes it possible to apply a standard spherical model to relaxor ferroelectrics, which predicts the experimentally observed quadratic divergence of the universal part of the susceptibility above the critical temperature. This model is complementary to the so-called spherical random bond – random field model proposed by Blinc *et al.* [14,15] to explain the NMR data and the non-linearity of the total dielectric susceptibility in relaxors.

In comparison with PMN, the crystal structure and polar order of PZN appear to be quite different. Earlier works reported that a phase transition from cubic to a rhombohedral phase took place upon cooling, which was associated with the maximum of dielectric constant occurring at $T_m$ around 410 K.[16-18] At room temperature weakly birefringent domains with extinction directions along $<110>_{cub}$ were observed on a $(001)_{cub}$–cut PZN platelet, which seems to confirm the rhombohedral symmetry. The value of birefringence decreases gradually upon heating but more sharply around 390 K before vanishing at T = 413 K.[19] Recently, Lebon *et al.* [20] reported that the cubic-to-rhombohedral phase transition in PZN is diffuse and spreads over the temperature range between 385 and 325 K with a full establishment of the rhombohedral phase below 325 K. This phase consists of domains of mesoscopic (60-70 nm) size. Application of a dc field along <111> transforms the polydomain state into a rhombohedral quasi-monodomain state. High-energy x-ray diffraction studies on PZN crystals by Xu *et al.* [21] revealed Bragg peaks resembling a tetragonal (or pseudo-cubic) symmetry (X-phase) for the bulk crystal, the nature of which is yet to be clarified. On the other hand, neutron scattering results showed the onset of diffuse scattering at the Burns temperature, and a softening of the optical mode at the critical temperature, analogous to PMN.[22]

Despite recent intense work, the nature of phase transitions and dielectric relaxation in relaxors has not been thoroughly understood. In this work, we have studied the structural transformation in PZN crystals by synchrotron x-ray diffraction and by analyzing the dielectric properties as a function of temperature and frequency at zero-field and under a dc field. .

## II. EXPERIMENT

PZN single crystals were grown by spontaneous nucleation from high temperature solution according to the method and conditions described in Ref. 23. A crystal plate of a triangle shape (4 mm in edge and 330 μm thick) was cut with large surfaces parallel to the $(111)_{cub}$ plane. The $(111)_{cub}$ faces were mirror polished using a series of diamond pastes down to 3 μm. For the dielectric measurements, the sample was sputtered with gold layers on the $(111)_{cub}$ faces in a central area of 1.5x1.5 $mm^2$. Two gold wires were attached to the electrodes using silver paste. For the poling of the sample, an electric field of 20 kV/cm was applied at room temperature and kept on for a half-hour. The crystal was then short-circuited to remove possible space charges injected.

X-ray diffraction experiments were carried out on the unpoled crystal using the X22A beamline (32 keV, $\lambda \approx 0.38$Å, with a penetration depth of about 30 µm at normal incident) from the National Synchrotron Light Source (NSLS) at the Brookhaven National Laboratories. The beamline is equipped with a four-circle Huber diffractometer, with Si (220) and Si (111) analyzer-crystals mounted in the diffraction path. The diffraction data were collected at temperature range between 420 K and 25 K upon cooling. The accuracy of the temperature measurement was within ± 5 K and the temperature stability within ± 2 K. The $\theta$–$2\theta$ scans were performed over selected angular ranges centered about the (200), (220) and, in particular, (222) cubic reflections. Least-squares method was used for fitting the diffraction line profiles to different shape function. In addition to the Gaussian function that describes the proper Bragg peaks at X22A, the Lorentz, pseudo-Voigt and Pearson VII functions were also tested to choose the best one.

Dielectric spectroscopic measurements were performed by means of a computer-controlled system consisting of a Solartron–1260 Impedance Analyzer and a Solartron–1296 Dielectric Interface, at various frequencies (100 Hz - 1 MHz) in the temperature interval between 310 K and 620 K. The measurements were carried out under various conditions: 1) Zero-field-cooling (ZFC); 2) Field-cooling (FC) for unpoled crystal by applying a dc bias field; 3) Zero-field-heating (ZFH) for prepoled crystal.

## III. RESULTS AND ANALYSIS

### A. Structural Transformation

Preliminary synchrotron x-ray diffraction experiments were undertaken on a pressed PZN powder sample obtained by crashing small single crystals at X22A in the Bragg-Brentano geometry. The data collected for several characteristic reflections show a single peak for $(200)_{cub}$, a double peak for $(220)_{cub}$ and a double peak for $(222)_{cub}$ reflections, which indicate a non-cubic and very likely a rhombohedral symmetry.

In the $(111)_{cub}$ crystal, the diffraction data around the $(222)_{cub}$ Bragg reflection were carefully measured as a function of temperature. The results obtained at some selected temperatures are shown in Fig. 1 (all diffraction data are intensity normalized by the peak value at each temperature, *i.e.* $I / I_{max}$). The peak at 415 K appears sharp and symmetric with an instrument solution-limited width, indicating the cubic structure (in accord with all previously published results) and the excellent quality of the crystal. On the other hand, the bottom of the peak is slightly broadened, more pronounced at the lower angle side. To visualize this fact better, the profile is shown separately in semilogarithmic scale in Fig. 2 (a). Such kind of basis broadening of x-ray and neutron diffraction peaks in PMN-based relaxors is usually related to the scattering by polar nanoregions.[24,25] In PZN this effect was also reported in the high-temperature phase, but only for the neutron scattering.[22] The full profile can be well fitted by the sum of Gauss and Lorenz function representing the Bragg and diffuse scattering contribution, respectively. Note that these two functions are centered at slightly different ? values, and the fitting are disturbed at the angles far from the center.

The spectrum at 390 K is almost identical to that of 415 K. But at lower temperatures the distinctive shoulders begin to appear from the both sides of the peak. Upon further cooling, these shoulders become more and more significant and the intensity of the major peak decreases (Fig. 3) indicating that some regions of the crystal undergo a structural distortion. At low temperatures two peaks expected for rhombohedral phase are clearly visible, but surprisingly, besides these two peaks, the significant shoulders still remain. This means that an additional phase (or phases) not noticed in the previous investigations exist in PZN crystal alongside with the rhombohedral phase.

In the temperature interval of 50 - 325 K the full line profile can be well fitted as a sum of five overlapping Gauss functions. Fig. 2 (b) demonstrates the fit at 50 K as an example. Two central contributions (i.e. those that give rise to the maximums on the diffraction profile) can be assigned to the rhombohedral phase [$(222)_R$ and $(-222)_R$ reflections correspond to low-angle and high-angle maximums, respectively]. The remaining three contributions are related to the other phase of lower (probably a monoclinic or triclinic symmetry).

The summarized intensity of the peaks related to the new phase (calculated as the sum of intensities of the corresponding fitted peaks) accounts for about 40% of the total intensity of all peaks, which means that this low-symmetry phase exists at a significant concentration. However, the magnitudes of the peaks of this phase are comparatively small for two reasons: i) the intensity is distributed over more than three peaks , and ii) the peaks are wider, e.g. at 50 K the values of the width at half maximum (FWHM) are about $0.07^o$ for the two most intense peaks of the new phase, which is much larger than the $(-222)_R$ peak ($0.024^o$) and the $(222)_R$ peak ($0.046^o$). The small magnitude of those peaks explains why they can be reliably detected only with the help of synchrotron x-ray diffraction having both the intensity and resolution much higher than the conventional x-ray technique.

Note that the central peak, which is single at high temperatures, remains the dominant peak throughout the cooling down to 50 K. Fig. 3 shows the variations of the position (in 2θ), FWHM and intensity of this peak as a function of temperature. It can be seen that the FWHM and the angle of peak position increase upon cooling, first very slowly but much more quickly starting from about 350 K. A sharp drop of peak intensity is also observed between 325 and 370 K, i.e. in the temperature range where the shoulders around the major peak become very pronounced. The line profile cannot be unambiguously fitted in this temperature range. All these features provide the evidence for the phase transition. Our results are consistent with the recent study of the (333) and (005) lines of PZN,[20] in that this phase transition is indeed diffused, i.e. in the temperature interval of 325-390 K the cubic phase transforms progressively into the domains of ferroelectric phase so that the different phases coexist in this interval. Due to the fact that the lattice plane spacings in the $[111]_{cub}$ direction are very close in the cubic and rhombohedral phases (*i.e.* the unit cell changes during the transition in such a way that its dimension in one of the $\langle 111\rangle_{cub}$ direction remains unchanged),[20] the rhombohedral $(\bar{2}22)_R$ and cubic $(222)_C$ reflections are superimposed and cannot be resolved, which is why only single peak composed of these two contributions can be observed in the temperature range of diffuse phase transition.

Fig. 4 presents the temperature dependences of the lattice parameters *a* and α and the unit cell volume calculated for the cubic and rhombohedral phases in PZN crystal. The anomaly around 350 K clearly indicates the phase transition. The value of α in the low-temperature R3m phase agrees satisfactorily with that reported for this phase in Ref. 20 and is approximately the same value as in the rhombohedral phase of normal perovskite ferroelectrics. Interestingly, the variation of the rhombohedral lattice constant *a* below the transition temperature (Fig. 4) shows the same trend as that of the PMN,[25,26] reflecting the relaxor behavior of PZN even in the low temperature phase(s). The width of the diffraction peaks below the phase transition temperature is much larger than in the high-temperature cubic phase. Fig 3 (b) illustrates the FWHM for the major peak. The widths of other peaks are even larger. This effect is usually explained by the small size of ferroelectric domains. The other, probably more important, reason for this in PZN is the internal elastic microstrains caused by the coexistence of different ferroelectric phases. The spontaneous deformations of the parts of the crystal containing different phases are different, which leads to internal stresses and strains. Spontaneous deformation usually increases with decreasing temperature, which is confirmed in our case by the increase of the rhombohedral angle in Fig. 4. As a result, FWHM also increases with decreasing temperature (Fig 3). Additional line broadening can also arise from a dispersion in lattice parameters, which depends on the distance from the crystal surface.[21]

Using the Scherrer equation the size of ferroelectric domains was estimated from the difference between the instrumental FHWM and the FHWM observed at 300 K (i.e at a comparatively high temperature, where the broadening related to internal strains is not very large). For the rhombohedral phase we derive a size of ~70 nm and ~600 nm from $(222)_R$ and $(\bar{2}22)_R$ peaks, respectively. This means that the domain size in one of the directions (perpendicular to the spontaneous polarization vector) is much larger than in the other directions, i.e. the domains have the laminar form. Note that in Ref. 20 all the dimensions of rhombohedral domains in PZN were estimated to be the same size (60-70 nm). This discrepancy can be explained by a longer x-ray wavelength used in that work leading to smaller penetration depth (the structure of PZN is known to depend on the distance from crystal surface[21]). The domain size of the new phase is estimated to be in the range of 40 – 120 nm, depending on the direction.

## B. Dielectric Properties

Figures 5(a) and 5(b) give the variations of the real part of dielectric permittivity $\varepsilon'$ and the dissipation factor tanδ as a function of temperature at various frequencies, measured upon zero electric field cooling (ZFC) for the $(111)_{cub}$ PZN crystal. The strong frequency dispersion of the dielectric constant with the temperature of the maximum, $T_m$, increasing at higher frequencies, indicates the typical relaxor ferroelectric behavior. The temperature $T_m$ varies from 417 K at 100 Hz to 429 K at 1 MHz. As in other relaxors, the frequency ($f$) dependence of $T_m$ can be fitted with the Vogel-Fulcher relation:

$$f = f_0 \exp[-E_a/(T_m - T_0)] ,$$

where $f_0$, $E_a$, and $T_0$ are the parameters, as shown in Fig. 5(c). The same fit, but with a different set of parameters can be applied to the frequency and temperature dependence of the imaginary permittivity maximum (not shown). The best-fit results are presented in Table 1. It is worth noting that below $T_m$ no evidence of the structural phase transition, which was detected between 390 and 325 K in the above mentioned x-ray diffraction experiments, can be observed in the dielectric properties.

Figure 6(a) shows the temperature dependence of the real permittivity of the unpoled crystal measured at various frequencies upon cooling under a dc bias field of 1.2 kV/cm (FC). The strong dispersion due to relaxor relaxation around $T_m$ remains. However, at $T_C \approx 390$ K, the dielectric constant undergoes a discontinuous change in slope with the values at different frequencies merging together and dropping sharply. Below $T_C$ the frequency dispersion is attenuated dramatically. The dielectric relaxation around $T_m$ can also be fitted with the Vogel-Fulcher relation for both the real and imaginary parts of permittivity, with the fitting parameters given in Table 1.

Fig. 6(b) presents the temperature and frequency dependences of the dielectric constant of the PZN crystal prepoled at room temperature (at 20 kV/cm), which were measured upon zero-field-heating (ZFH after poling). In the low temperature range, the permittivity is almost non-dispersive. Upon further heating, a sharp peak of dielectric constant occurs at $T_C$ = 388 K. Above $T_C$, the strong dielectric dispersion, characteristic of relaxor relaxation, reappears, suggesting that the PZN crystal reenters the relaxor state. The transition temperature $T_C$ does not depend on frequency, as opposed to the behavior of $T_m$. The frequency dependence of the latter can also be well fitted into the Vogel-Fulcher law with the fitting constants provided in Table 1.

It is interesting to note that i) above $T_C$ the electric field almost has no effect on the dielectric relaxation behavior around $T_m$, which can be fitted into the Vogel-Fulcher relation with fitting parameters only slightly different from those of ZFC, ii) the sharp anomalies of the dielectric constant upon ZFH after poling and upon FC are observed at approximately the same temperature as $T_C \approx$ 390, where the diffuse phase transition begins upon ZFC, as revealed by synchrotron x-ray diffraction experiments.

## IV. DISCUSSION

Let us discuss the structure and properties of the PZN crystals studied in this work by comparing them with the well-documented structure and properties of the prototypical relaxor ferroelectric PMN (see e.g. Ref. 1). The temperature and frequency dependences of the dielectric permittivity look very similar in both crystals, with observed broad and high $e(T)$ peak and strong dispersion causing the Vogel-Fulcher type shift of $T_m$ with frequency (see Fig. 5). In both crystals additional anomalies in the temperature dependences of permittivity and losses that are initially absent at zero field, can be induced by applying a strong enough electric field (see Fig. 6). In both cases, the high-temperature slope of the diffused permittivity peak can be scaled by the Lorenz-type quadratic function with the close values of diffuseness parameter ($d$ = 28 K for PZN and $d$ = 41 K for PMN), as described in detail in Ref. 27 (the date used for scaling came from the same PZN crystal as in this work). The structures of PMN and PZN at high temperatures (around $T_m$ and above) are also similar. It is usually believed that in PMN the structure is macroscopically cubic with the nanometric inclusions of polar order. In PZN the existence of PNRs have recently been deduced from neutron scattering experiments in Ref. 22 and confirmed in the present work by the observed broadening of the basis of the diffraction peak. On the other hand, the low-temperature structures of these two materials are quite different. In PMN the x-ray and neutron diffraction investigations do not indicate any macroscopic distortion of the cubic lattice. PZN, on the contrary, exhibits the reduction of symmetry below about 350 K where we observed the splitting of (222) lines. Two central contributions (peaks) can be attributed to the rhombohedral phase which was also observed in the previous investigations (e.g. in Ref. 20). In addition, we have revealed the presence of another low-symmetry phase with a significant concentration which was not reported before [the present data do not allow us to determine the symmetry and the type (ferroelectric or antiferroelectric) of this new phase; such an investigation is underway]. We also confirmed that the phase transition in PZN is diffused, i.e. the high-temperature cubic phase and the low-temperature phases coexist in a temperature interval of several dozens of degrees.

Note that the studies of the PZN crystal with the help of neutron and high-energy (67 keV) x-ray diffraction revealed a different low-temperature phase (the so-called X-phase) but not the rhombohedral one.[21,22,28] The X-phase exhibits a nearly cubic unit cell with a slight tetragonal distortion. It was not observed in this work, nor was it other works in which low-energy x-ray was used. This discrepancy can be explained[21] by the small penetration capability of low-energy radiation, so that it probes only the parts of the crystal not far from the surface ("skin"). X-phase seems to be located in the bulk and can be detected only by the high-energy radiation. As the phase content depends on the distance from the crystal surface (X-phase inside, "normal" phases at the surface), one can suspect that the rhombohedral phase and the additional low-symmetry phase discovered in this work are also separated in space. Further experiments are needed to determine if these two phases are mixed homogeneously or exist separately.

The important point to underline here is that, the PNR-related diffuse scattering giving rise to the tails around the sharp $(222)_{cub}$ Bragg peak at high temperatures has been observed in our x-ray diffraction experiments. This means that PNRs exist not only in the crystal bulk, transforming to the X-phase upon cooling, but also in the "skin" of the crystal which undergoes the transition into the phases with normal ferroelectric distortion.

It is interesting to compare the behavior of PZN with those of other perovskite materials which exhibit a spontaneous relaxor to normal ferroelectric phase transition [e.g. $Pb(Fe_{1/2}Nb_{1/2})O_3$,[29] $Pb(Sc_{1/2}Nb_{1/2})O_3$,[30] or $Pb(Mg_{1/3}Nb_{2/3})O_3$ – $PbTiO_3$ solid solutions with high concentration of $PbTiO_3$]. In these crystals the characteristic diffuse $\varepsilon$ ($T$) peak exhibiting the Vogel-Fulcher frequency dependence is accompanied by a dielectric anomaly at several degrees below $T_m$. This anomaly is related to the spontaneous (i. e. without external field) transition to the ferroelectric phase upon cooling and it can be very sharp. Below the phase transition temperature well-defined ferroelectric phase exist with macroscopic domains of ~1 μm in size.

One can see that the behavior of the PZN crystals sits in an intermediate position between the behavior of prototypical relaxor PMN and that of the crystals with the sharp spontaneous relaxor to normal ferroelectric phase transition. The spontaneous transition to the ferroelectric phase is observed in PZN, but this transition is diffused and thus it is not associated with the sharp dielectric anomalies. The size of the ferroelectric domains is considerably smaller than the size of normal ferroelectric domains, but larger than the size of polar nanoregions in PMN.

To interpret our results we apply the kinetic model of phase transitions, which is developed to describe the diffuse and sharp phase transitions in compositionally disordered crystals.[31] According to this model the PNRs begin to appear within the paraelectric phase at $T_d >> T_{0m}$ ($T_{0m}$ is the average temperature of ferroelectric phase transition) as a result of local "phase transitions" caused by compositional inhomogeneities in disordered crystal (the nature of these inhomogeneities and the peculiarities of polar order inside the PNRs are discussed in Refs. 32 and 13). The equilibrium size and number of PNRs gradually increase during cooling. At a certain lower temperature $T_C$ the PNRs become metastable and their sudden thermally-activated growth is possible (similar to the isothermal growth of the nuclei of a new phase in the case of the normal first-order phase transition). The model parameter $r_c$ (which is directly proportional to $T_{0m}$ and inversely proportional to the diffuseness of the phase transition) determines the fraction of crystal bulk filled with PNRs at $T = T_C$. If $r_c$ is comparatively small, the concentration of PNRs at $T_C$ is large and the thermally-activated growth of any PNR is limited by the neighboring PNRs as well as by the areas having a lower local Curie temperature. To form a large polar domain in some region,

PNRs have to merge, i.e. the directions of their dipole moments have to change to be the same for all PNRs in this region. But the reorientation of all PNRs appears to be impossible because at least some of them are frozen. This freezing can be due to one of the following reasons: i) the temperature is too low to activate the PNRs so as to overcome the potential barrier between the states with different directions of PNR dipole moment, ii) dipolar glass state is formed in which the directions of PNR moments are fixed by the frustrated interactions between them, iii) PNR moments are pinned by local random electric and/or elastic fields. Consequently the size and number of PNRs remain almost unchanged when the crystal passes through $T_C$. As a result, the long-range polar order characteristic of ferroelectric state cannot develop and thereby no noticeable anomalies of structural parameters and dielectric (and other physical) properties can be observed. This scenario seems to be valid for PMN in which PNRs are commonly believed to exist at all temperatures below $T_d$ and all the hree reasons mentioned above for their freezing can be expected. In PZN the $r_c$ parameter is larger due to a higher $T_{0m}$ and a smaller (see above) phase transition diffuseness $d$. A lager $r_c$ means that at $T_C$ the concentration of PNRs is smaller (the distance between them are rather large) and they have room to grow up to mesoscopic sizes large enough to be detected by X-ray and neutron diffractions. The higher transition temperature in PZN probably facilitates the formation of the larger polar regions in another way. At a higher temperature the dipole moments of some PNRs can be reoriented by thermal motion. Consequently the growth of PNRs at $T_C$ is accompanied by the reorientation of some of the neighboring PNRs so that several PNRs can merge to form larger ferroelectric domains (the merging decreases the energy related to the domain walls). Upon further cooling below $T_C$ the process of domain formation goes on because of the increase in ferroelectric distortion and the transformation to the ferroelectric phase of the regions with reduced local Curie temperature. This process has been revealed in the diffraction experiments. The corresponding anomalies (Fig. 3) are observed not at well-defined temperature, but smeared over a wide temperature interval.

Materials with a sharp transition from relaxor to normal ferroelectric state are usually characterized by a small diffuseness parameter $d$ and consequently a large $r_c$ parameter. As a result, the concentration of PNRs is small at $T_C$ and they are free to grow into macroscopic ferroelectric domains.

The dielectric behavior of relaxors is also determined by the kinetics of the formation and evolution of the PNRs and ferroelectric domains. This is because the dielectric response of relaxors in the temperature range around $T_m$ arises mainly from the relaxation of PNRs and their boundaries, rather than from the non-relaxation ionic polarization related to the relative displacement of the positive and negative sublattices, as in the case of normal displacive ferroelectrics (see e.g. Ref. 33 for more detailed discussion). In the relaxors that do not undergo a transition into the ferroelectric phase upon cooling (e.g. in PMN) the temperature evolution of PNRs occurs without abrupt changes in their size and concentration and consequently there are no sharp anomalies in the temperature dependences of permittivity. The relaxors with a sharp spontaneous phase transition into the normal ferroelectric phase show an abrupt drop of dielectric constant at $T_C$ due to the transformation of PNRs into macroscopic ferroelectric domains at this temperature. In the intermediate case of PZN, the transformation of PNRs into ferroelectric domains takes place gradually so that the dielectric permittivity changes without noticeable anomalies. At temperatures above the diffuse phase transition (including $T_m$) the permittivity is determined by the relaxation of PNRs. At temperatures below the transition the main contribution to the permittivity comes from the relaxation of the walls between mesoscopic domains. In the temperature range of diffuse transition both mechanisms apply.

The electric field applied on the PZN crystal upon cooling is able to reorient PNRs, so that all of them have the same (or almost the same) orientations of dipole moments and at $T_C$, where the process of intensive growth of PNRs begins, they can easily merge to form macroscopic ferroelectric domains. The number of relaxing elements (e.g. domain walls and boundaries between different phases) that are able to contribute to the dielectric constant decreases rapidly during this process, leading to the distinct dielectric anomaly at $T_C$. Below $T_C$, the dielectric dispersion is almost suppressed [Fig. 6 (a)]. Upon heating of the poled crystal, the phase transition occurs at $T_C$ accompanied by the sharp dielectric peak [Fig. 6(b)] indicating that the crystal transforms back to the same relaxor state as it was in the zero-field experiments, with the presence of PNRs. As a result, the Vogel-Fulcher parameters remain almost unchanged (see Table 1). However, in contrast to the zero-field experiments, the orientations of PNRs are no longer random. Instead, the PNRs subsystem is poled (or partially poled) so that the magnitude of the permittivity is different (smaller).

It should be pointed out that the phenomenological kinetic model discussed above does not take into account the possible symmetry of ferroelectric phase. For such consideration, the “soft polar nanoregions” model[13] can be applied. It is argued in this model that due to the randomness of microscopic forces responsible for the onset of spontaneous polarization, each PNR consists of differently polarized unit cells. This model implies that the local symmetry of the structure inside the PNR can randomly vary in space and probably with time. We believe that different local symmetry prevails in different PNRs so that they can develop upon cooling into the ferroelectric domains of different symmetry. As a result at least two phases are present in PZN at low temperatures.

## IV. CONCLUSIONS

We have shown in this study that the $Pb(Zn_{1/3}Nb_{2/3})O_3$ crystal is a unique example of relaxor in which, in contrast to classical relaxor PMN, the spontaneous (i.e. without external electric field) ferroelectric phase transition occurs, but, in contrast to some other relaxors exhibiting sharp spontaneous transition to a ferroelectric phase [e.g. $Pb(Fe_{1/2}Nb_{1/2})O_3$], this transition is diffuse. As detected by high-resolution synchrotron x-ray diffraction in the absence of an electric field, PZN crystal undergoes a diffuse structural transformation from the high-temperature state, which is macroscopically cubic and contains polar nanoregions typical of relaxors, to the low-temperature state composed of the mesoscopic (40 - 600 nm) domains of the rhombohedral ferroelectric phase and a second phase with lower symmetry. On cooling the domains of these low-temperature phases begin to appear at $T_C$ » 390 K and grow progressively at the expense of the cubic phase. Below $T$ » 325 K the cubic phase is no longer observable. The so-called X-phase recently discovered in the central (bulk) parts of PZN crystal with the help of high-energy x-ray and neutron diffraction, was not observed in the present work because the x-ray energy used (32 keV) was not high enough.

The dielectric properties show typical relaxor ferroelectric behavior with the broad and dispersive peak of dielectric constant at $T_m \approx 415$ K > $T_C$, which can be fitted into the Vogel-Fulcher relation, while no clear anomalies in dielectric properties can be associated with the structural phase transformation at $\sim T_C$. Application of an electric field (1.2 kV/cm) upon cooling induces a comparatively sharp phase transition at $T_C \approx 390$ K with the establishment of the ferroelectric phase with macroscopic domains, as revealed by the anomaly in the temperature dependence of dielectric constant at $T_C$ and the disappearance of significant dielectric dispersion below $T_C$. The state induced upon field-cooling collapses under ZFH at $T_C$ in the form of a sharp phase transition with the breaking down of the macro polar domains back into the relaxor state. The relaxor behavior is fully recovered with typical relaxor dielectric relaxation around $T_m \approx 415$ K and the same fitting parameters to the Vogel-Fulcher relation.

The structural behavior, dielectric properties and phase transition of the PZN crystal are discussed in the light of the kinetic model of phase transitions in disordered crystals and the model of "soft nanoregions" in relaxors.

ACKNOWLEDGMENTS

The authors thank W. Chen for help in crystal preparation, and G. Xu, Z. Zhong and C. Stock for useful discussion. This research was supported by the Natural Sciences and Engineering Research Council of Canada (NSERC), the U.S. Office of Naval Research (Grant No. N00014-99-1-0738) and the U.S. DOE (Contract No. DE-AC02-98CH10886).

**Figure Captions:**

Figure 1. Diffraction pattern around the $(222)_{cub}$ peak for PZN crystal at selected temperatures between 50 K and 415 K (with intensity normalized by $I / I_{max.}$).

Figure 2. Fitting of the diffraction pattern around the $(222)_{cub}$ peak for PZN crystal at: (a) 415 K and (b) 50 K. Circles represent experiment data; Gaussians and Lorentzian used for fitting are represented by solid and broken lines, respectively.

Figure 3. Variations of (a) two-theta (2θ) values, (b) full-width-at-half-maximum (FWHM), and (c) integrated intensity of the major $(222)_{cub}$ peak as a function of temperature for PZN crystal.

Figure 4. Variations of lattice parameters, *a* and α, and the unit cell volume, as a function of temperature for the rhombohedral and cubic phases of PZN crystal

Figure 5. Variations of, (a): the real part of dielectric permittivity, and (b): the dissipation factor, as a function of temperature for PZN crystal upon cooling at zero-field (ZFC) measured at different frequencies. (c): Frequency dependences of the temperature ($T_m$) of maximum permittivities (circles for real part and stars for imaginary part) and the fitting (solid line) to the Vogel-Fulcher relation.

Figure 6. Variation of the real part of dielectric permittivity measured at different frequencies as a function of temperature for PZN crystal upon: (a) cooling under a field of 1.2 kV/cm, and (b) heating after poling at room temperature.

Table 1. Fitting parameters of the Vogel –Fulcher relation for the temperatures of maximum real and imaginary permittivities obtained under different conditions

| | ZFC | | FC | | ZFH after poling | |
|---|---|---|---|---|---|---|
| | ε’ | ε” | ε’ | ε” | ε’ | ε” |
| $f_0$ (Hz) | 3e+10 | 8e+10 | 5e+10 | 2e+8 | 3e+9 | 5e+11 |
| $E_a$ (K) | 263 | 488 | 332 | 165 | 206 | 526 |
| $T_0$ (K) | 403 | 380 | 393 | 392 | 401 | 377 |

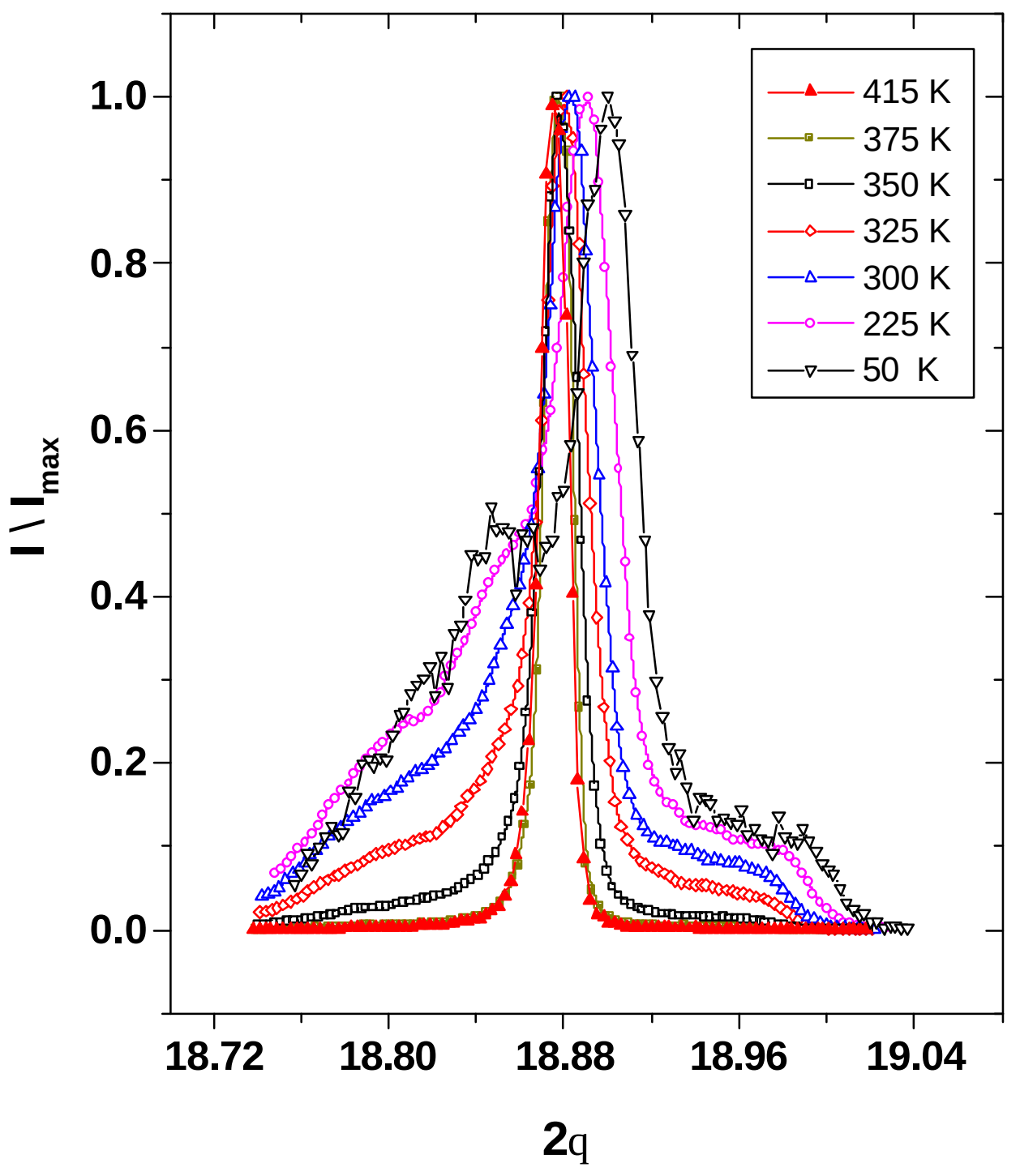

Fig. 1

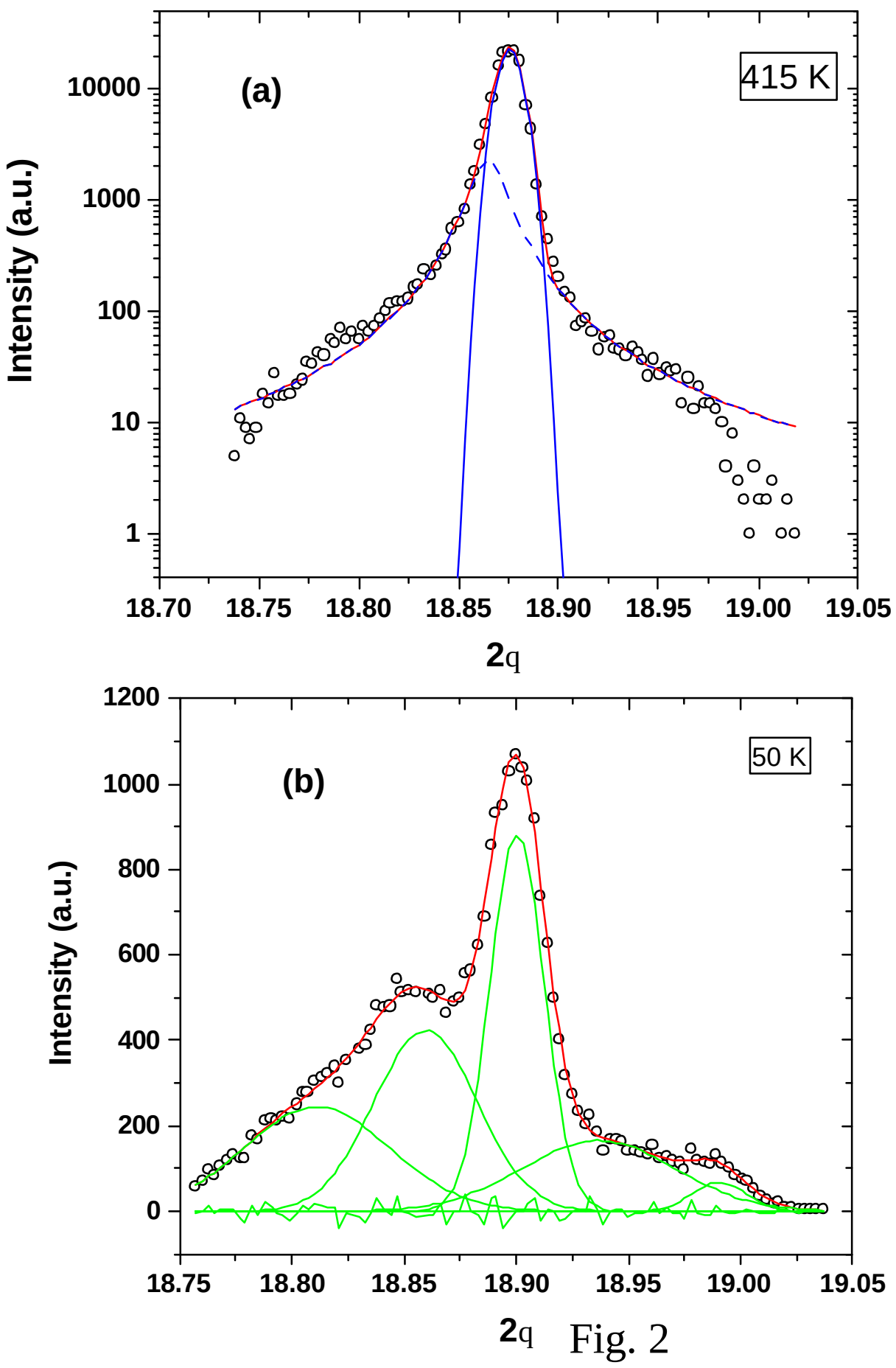

Fig. 2

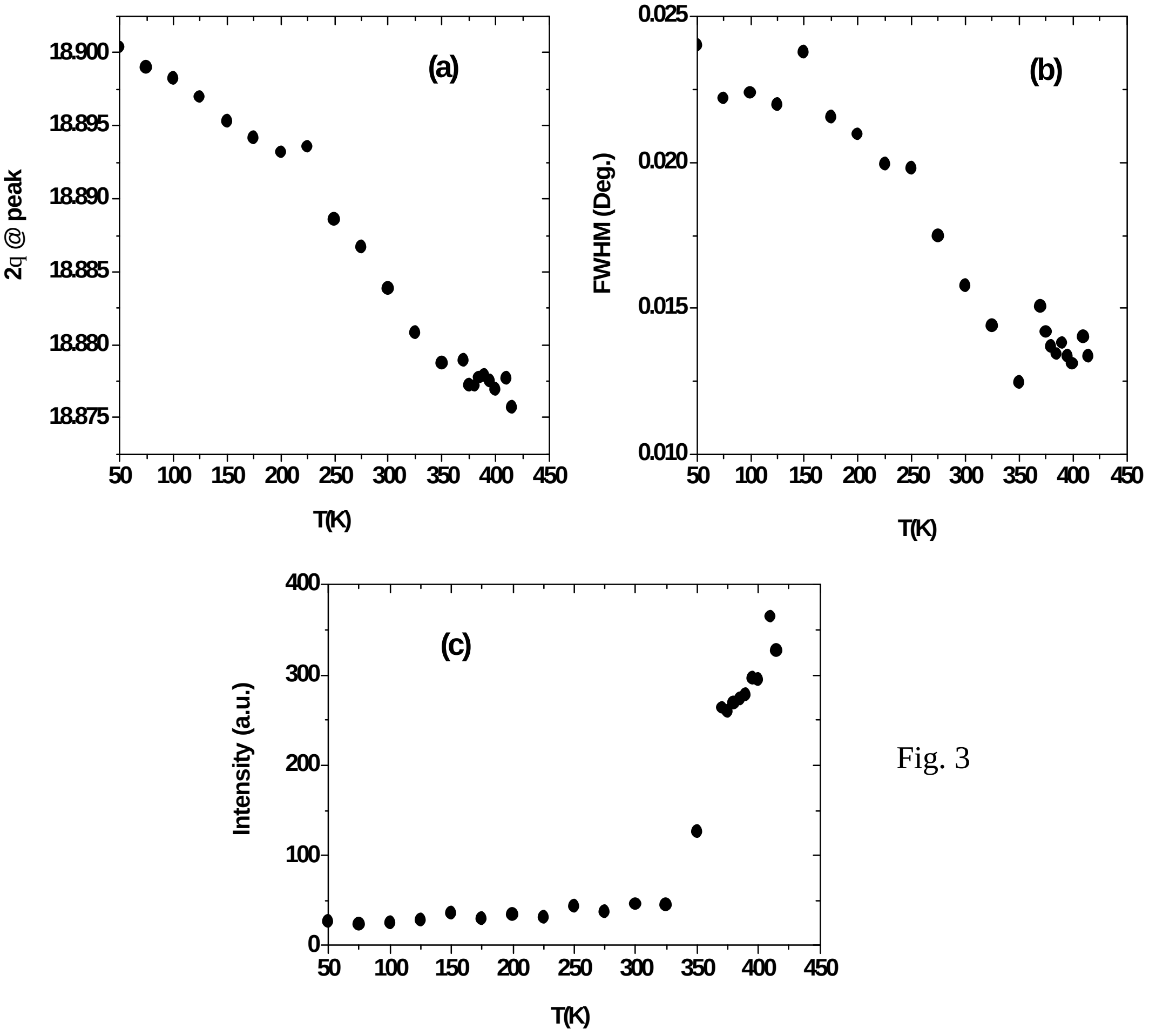

Fig. 3

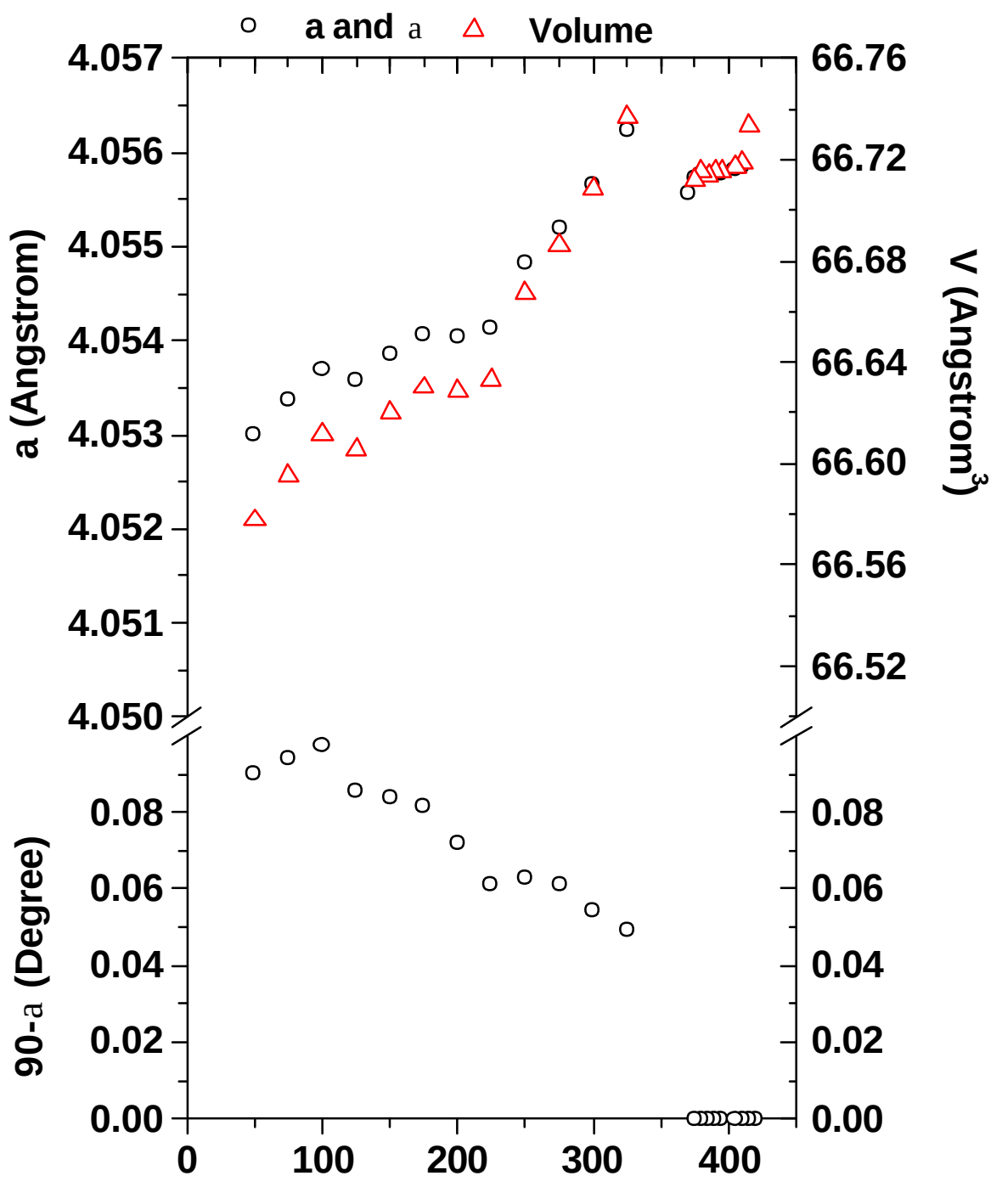

Fig. 4

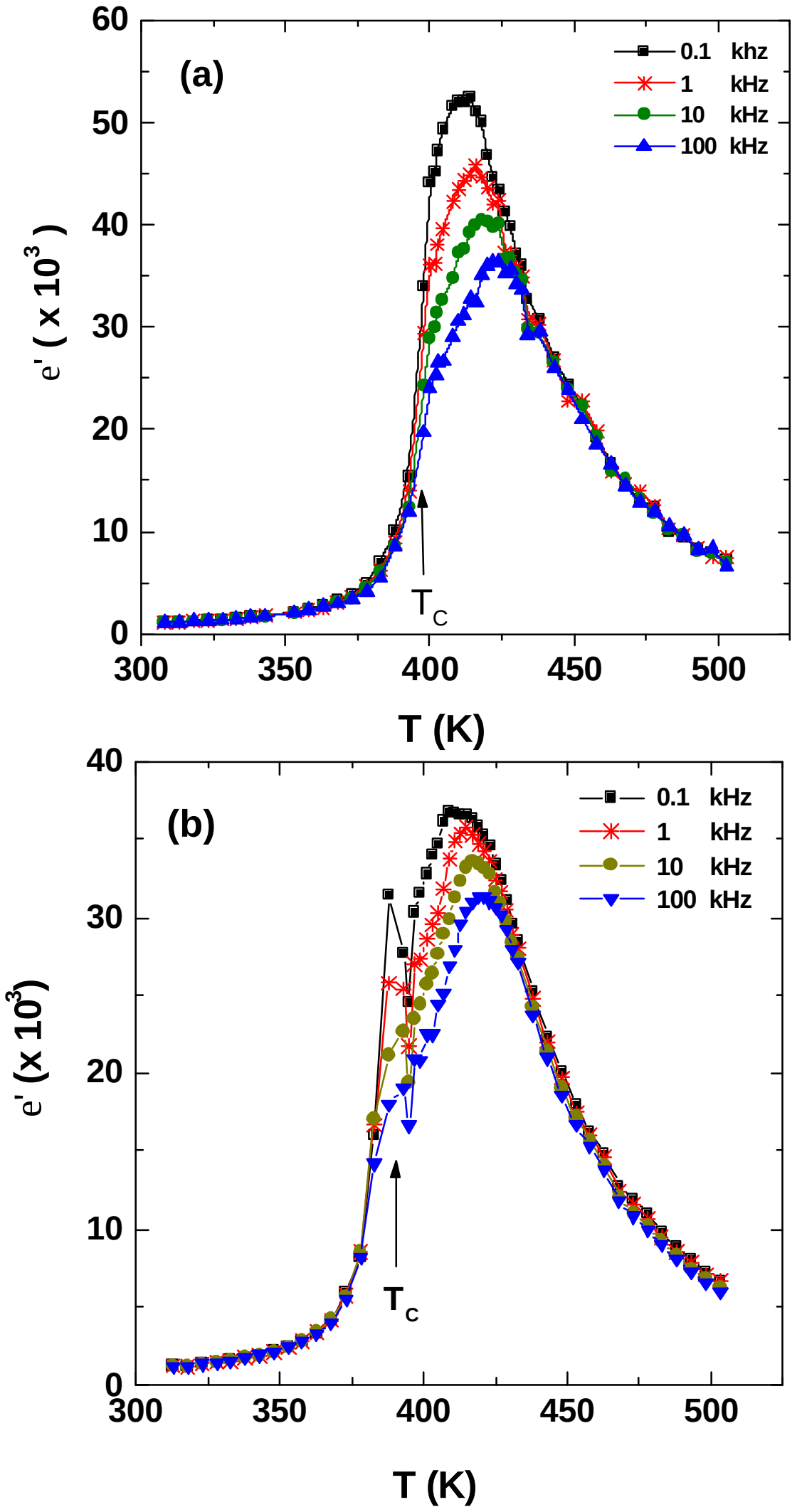

Fig. 6

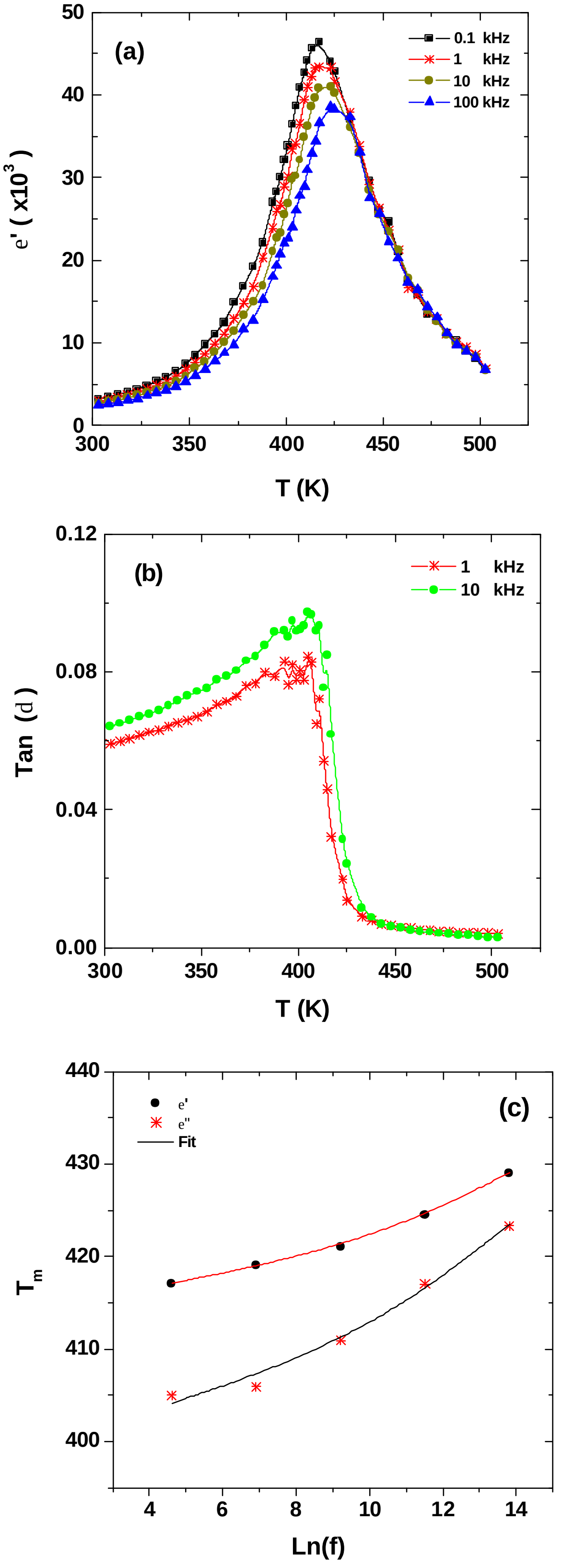

Fig. 5